# High-pressure NMR spectroscopy in studies of the conformational composition of small molecules of ibuprofen in supercritical carbon dioxide

*Ilya Khodov,[a] Alexey Dyshin,[a] Sergey Efimov,[b] Dmitry Ivlev,[a] Michael Kiselev,[a]*

[a] *Laboratory of NMR-spectroscopy and numerical investigation methods of liquids G.A. Krestov Institute of Solution Chemistry, Russian Academy of Sciences, Akademicheskaya str. 1, Ivanovo, 153045, Russian Federation*
[b] *Institute of Physics, Kazan Federal University, Kremlyovskaya str. 18, Kazan, 420008, Russian Federation*

ABSTRACT

An experimental approach in conducting NMR measurements at supercritical parameters of state is discussed. A novel design of the high-pressure NMR cell was developed which allowed eliminating the field inhomogeneity and, thus, increasing the sensitivity of the experiment at the supercritical state. Analysis of the MD simulations and NMR data showed that two conformers of ibuprofen dominate in the solution in supercritical $CO_2$ along the critical isochore 1.3 $\rho_{cr}(CO_2)$. Conformer populations calculated from MD simulations and from NMR spectra agree with each other

## 1. Introduction

Obtaining quantitative information on the conformational distribution of small biologically active molecules is important to predict their bioavailability and describing the nucleation of polymorphic forms. Screening of crystalline forms is carried out at initial stages of developing drug compounds in order to identify polymorphs and pseudopolymorphs [1,2], since only pure and stable crystalline forms are admitted to the market. Structure of the crystal lattice is defined by two main factors: molecular packing and conformation [3]. The case when various polymorphic modifications are built from different conformers is of special interest and is called conformational polymorphism. Investigation of conformational polymorphism, is justified by the fact that different polymorphic forms usually have different sublimation enthalpies, which define, to a great extent, their solubility [4]. Formation of polymorphic forms is a complex, multistep process, which depends on different thermodynamic and kinetic factors [5]. The most widely accepted theory of crystal formation from a saturated solution begins with appearing of pre-nucleation clusters, which then lead to formation of seeds and then crystals of a certain structure. It was shown in [6] that a saturated solution contains molecule clusters reflecting unit cells of all possible polymorphs. The Formation of a certain polymorph is dependent on different crystallization conditions such as supersaturation degree, solution type, and the presence of a cosolvent, temperature etc. )

Crystallization from supercritical fluids attracts much interest as an ecologically friendly replacement of organic solvents in obtaining new crystalline forms. Supercritical carbon dioxide is an alternative solvent to use in pharmacy due to its physical and chemical properties: it is nontoxic, inflammable, has relatively low values of critical pressure and temperature (73.8 bar, 31°C), and it is cheap. Simple decompression is enough to separate a solute from the solvent without the need of power-consuming drying and purification. Two methods of obtaining crystalline forms of drug compounds utilizing supercritical solvents can be mentioned: rapid expansion of supercritical solutions (RESS) and the method based on supercritical antisolvent (SAS) precipitation (the latter is used for poorly soluble in supercritical $CO_2$ compounds). In both cases, obtaining of polymorphic forms depends on the predominant molecular conformation in the solution at the supercritical parameters of state. Therefore, developing approaches to control the conformational ensembles of molecules in a supercritical solvent is very important. A significant progress in studies of conformational state of molecules of biologically active compounds has been achieved recently using NMR spectroscopy [7–16] and vibrational spectroscopy [17–23].. A correlation between the distribution of the conformers of ibuprofen in supercritical carbon dioxide and fraction of the polymorphs arising in supercritical sedimentation has been found with the aid of vibrational spectroscopy and MD simulations [20]. This example shows clearly that vibrational spectroscopy can give information on the conformational manifolds of small molecules which would be in a good agreement with MD simulations. However, results given by vibrational spectroscopy are indirect, because conformer populations are derived by decomposing spectral bands into elementary components, which cannot be obtained by an unambiguous mathematical procedure; hence, vibrational spectroscopy cannot yield direct information of the fraction of conformers. An alternative method is the NMR spectroscopy at high pressures, which has shown its usefulness in studies of molecular structure [24–26], host-guest interactions [27], metal–organic framework[28,29], ionic liquids [30–32], aqueous geochemistry [33–36] and operando studies of complex mixtures[37]. In addition, liquid state NMR spectroscopy is useful in studies of conformational exchange [38–42]. However, conducting NMR experiments under supercritical conditions requires a specialized high-pressure supercritical cell.

The aim of the present work is developing the method of determination of the conformational equilibrium of small molecules using high-pressure NMR spectroscopy on the example of ibuprofen in supercritical $CO_2$. This will facilitate investigations dedicated to the nucleation mechanism of conformational polymorphs at the molecular level, conducted in supercritical fluids.

## 2. Results and Discussion

### 2.1. High-pressure NMR spectroscopy

Appearing of high-pressure NMR spectroscopy for studying chemical properties of water and organic solvents was connected with development of a novel NMR probe head designed as a autoclave, in which the sample and electric circuits are placed under pressure in a metal chamber [43]. A number of interesting results were obtained with the aid of this probe [44,45], but it cannot compete with modern commercially available devices. On the other hand, lately, home-built probe head are significantly modernized and give good results in comparison with the first versions of the autoclave type probe head. For example, there is a specialized high-pressure liquids probe head developed by William Casey's group at UC Davis [46] and the high-pressure MAS NMR probe heads/rotors developed at Pacific Northwest National Laboratory by Eric Walter and David Hoyt [47]. Thus, an alternative solution was chosen for obtaining NMR spectra under high pressure: special high-pressure tubes were used in combination with standard probe heads, without the need to modify their construction. Special inserts with a small inner diameter (quartz capillaries) were used initially [48]. They allow achieving high pressures, but a decreased useful volume and decreased accordingly signal-to-noise level are a serious drawback. Later sapphire NMR tubes with a large active volume were brought into practice, which allowed us to increase significantly the sensitivity. However, attaching of these tubes to the high pressure source was a difficult task. First, the tubes were stuck to the pressure valve by glue [49]. Use of epoxy compound as a glue is not very reliable at high pressure and hence dangerous for the magnet and probe of NMR spectrometer. Later, using epoxy compounds was limited to using flange connections. A more reliable strategy of sealing the cell was proposed in [50], which was based on a novel connection valve in a multistep high-pressure flange. High pressure ceramic NMR tubes were developed to replace sapphire tubes [51]; their advantages, however, are still not obvious.

A special sapphire cell modified for carrying out experiments in the supercritical state was employed. The cell allowing obtaining NMR spectra of compounds under the pressure up to 300 bar and temperature up to 90°C was turned from a single crystal of sapphire (Figure 1, pos. 3); it has the outer diameter of 5 mm, inner diameter of 3mm, and length of 87 mm. The tube was annealed to remove internal stress. The sapphire tube was purchased as part of a high-pressure NMR cell with an integrated valve designed for Bruker spectrometers (Daedalus Innovations, Aston, PA, USA).

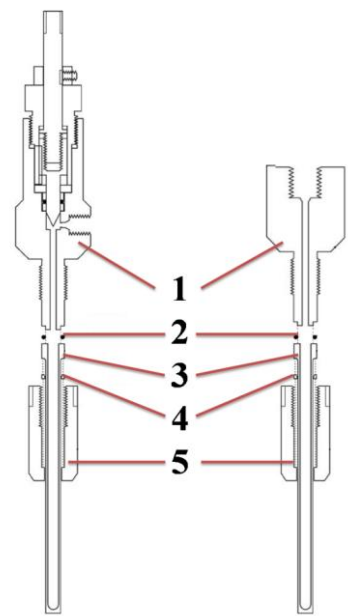

**Fig. 1 -Design of supercritical NMR cells manufactured by Daedalus Innovations, LLC (left) and after our modification (right). (1) Cell body (the left picture shows the body with the integrated high-pressure valve; the right picture, the cell body with gas inlet port for 1/16 in. tubing to which the high-pressure capillary is attached directly); (2) O-shaped seal (left – Viton O-Ring from Daedalus Innovations, LLC; right – MVQ 70 ShA O-Ring from Bohemia Seal, s.r.o., Czech Republic); (3) sapphire high-pressure tube; (4) tube seat (compensating Caprolone padding); (5) coupling ring.**

A few design drawbacks of the Daedalus cells were revealed during the measurements:

Metal body of the cell (Figure 1, pos. 1) exerts too strong influence on the magnetic field homogeneity;

Pressure in the cell cannot be controlled in real time, which is an important prerequisite for working at different parameters of state (temperature and pressure);

Unlucky design of the junction between the metal body and the tube itself led to appearing of cracks in the sapphire walls near the metal–sapphire contact place;

The sapphire tube decreases the sensitivity of the measurements due to a strong absorption of the signal by the walls.

To obtain correct results, it was necessary to eliminate or decrease the influence of the mentioned faults.

To diminish the influence of the metal high-pressure NMR cell body on the magnetic field homogeneity, heat-strengthened aluminium alloy was used to construct a homemade cell body. A screwed joint with a wedge-type seal was made to fasten a high-pressure capillary, which is used to fill the tube with supercritical carbon dioxide and to control its pressure at the same time. The metal body was made as small as possible in order to reduce the influence on the magnetic field and decrease the dead volume of the cell. The inlet capillary orientation was adjusted using polymer washers so that it was vertical and thus distorted the field insignificantly. The landing thread for the Viton seal (Figure 1, pos. 2) was made shorter. At the same time the diameter of the coupling ring (Figure 1, pos. 5), which pulls the sapphire tube to the metal body of the cell, was changed to decrease the degree of deformation of the sealant. This modification allowed improving the accuracy of the pressure control due to more correct operation of the sealing unit. The pressure was produced and controlled with the accuracy of ±0.5 bar using a high-pressure valve system and a standard laboratory model of hand-powered press (model 37-6-30) manufactured by HiP (High Pressure Equipment

Company, PA, USA). The NMR measurements at supercritical state parameters were tested on the example of ibuprofen, which has already been studied well in the supercritical solutions [52–56].

Ibuprofen belong to the class of nonsteroidal anti-inflammatory drugs (NSAID) of the 2-arylpropionic type, which are also known as "profens." They inhibit cyclooxygenase through interaction with the active centre of the enzyme [57].

Molecules of ibuprofen contain a benzene ring connecting two *para*-substituents, isobutyl groups and propionic acid. These substituents provide the molecule conformational flexibility, which leads to arising of multiple conformers [58]. Two polymorphic forms of the crystalline state of ibuprofen are known today [59]. The difference between the polymorphs is due to different orientations of the carboxyl group in the propionic acid moiety; knowledge of the orientation of this group may improve our understanding of the growth of crystals with different structures.

In addition, a fundamental knowledge of conformational features of biologically active molecules, involving NSAIDs, may shed light on their structure–activity relationships and thus elucidate their physiological activity. Conformational structure influences also the transport properties of a drug [60,61], its release from delivery systems [62], and hence its bioavailability. Experimental determination of the spatial structure and conformational state of pharmaceutical preparations still remains a challenging task.

### 2.2. Computer simulation

The molecular dynamics (MD) simulation was performed in this study to analyze the ibuprofen conformational manifold. The Gibbs free energy of the conformational transitions was calculated. Transitions between the conformers separated by high barriers are rare, which makes the simulation procedure quire complicated. To solve this problem, effective sampling procedures are used. One of the most efficient is metadynamics method which was used in the present study. The main idea of the method is to add an external potential consisting of Gaussian functions to provide the low-probability transitions [63,64]:

$$V_G\left(\vec{x}, t\right) = \sum_{t'=t_G, 2t_G, 3t_G, \dots} w \exp\left(-\frac{\left(\vec{x} - \vec{s}_{t'}\right)^2}{2\sigma\sigma^2}\right) \quad (1)$$

where $s_{t'} = S(x(t))$ is the value of the collective variable at time $t$; $V_G$ is the potential defined by the Gaussian height $w$, mean standard deviation $\sigma_S$, and frequency $t_G$ of adding the Gaussians into the total potential. These parameters influence the accuracy and efficacy of the description of the free energy surface

Metadynamics simulations were performed in the PLUMED software [65]. The calculation of the free energy surface was carried out by scanning over a pair of collective variables selected by us as dihedral angles associated with the inner rotation of the OH group of the ibuprofen molecule in step of 5 degrees.

The choice of the potential of intermolecular interaction is crucial for the correct carrying out of a computer simulation. In our previous work [66], a comparative analysis of the results of MD simulation using several types of interaction potentials and quantum-mechanical calculations was performed. Based on this analysis, the GAFF potential was selected for the ibuprofen molecule [67]. MD simulation of ibuprofen in supercritical $CO_2$ was carry out by using the NAMD program package [68] in NVT ensemble at the temperature of 50°C and the density of $CO_2$ corresponding to the pressure of 130 bar. Density values were taken from the NIST database [69].. Packmol program package [70] was applied to create the initial configuration of one ibuprofen molecule in 216 $CO_2$ molecules. We used the Zhang [71] potential model for $CO_2$ molecule.

Geometry parameters and atomic charges of the ibuprofen molecules were obtained from a quantum-chemical calculation using the functional B3LYP, basis cc-pVTZ [66].

The following formula was applied to calculate the probability density of two investigated dihedral angles being in certain ranges based on the free energy:

$$p(\Phi, \Psi) = \frac{\exp(-F(\Phi, \Psi)/RT)}{\int d\Phi \int d\Psi \, \exp(-F(\Phi, \Psi)/RT)} \quad (2)$$

Numerical values of the probability of formation of the conformers were calculated from the following expression [66]:

$$P = \int_{\Phi_1}^{\Phi_2} d\Phi \int_{\Psi_1}^{\Psi_2} d\Psi \cdot p(\Phi, \Psi) \quad (3)$$

where $\Phi$ and $\Psi$ are the dihedral angles, and integration is performed in the ranges from $\Phi_1$ to $\Phi_2$ and from $\Psi_1$ to $\Psi_2$, respectively; $p(\Phi,\Psi)$ is the probability density of arising of a given configuration.

### 2.3. MD simulations and calculation of the conformers

It was shown earlier that a set of dihedral angles (see Figure 2) is a representative set of collective variables for scanning the conformational space of ibuprofen [66]. The metadynamics method yielded a 2D surface of the free energy of the ibuprofen conformers with respect to the collective variables (see Figure 3). This 2D surface indicates the presence of four conformers combined in groups of two (A, B and C, D); the barrier between the members inside each group is smaller than the level of thermal fluctuations (see Figure 3). Thus, two conformers appear in practice, which are called (A+B) and (C+D) in the figure 3.

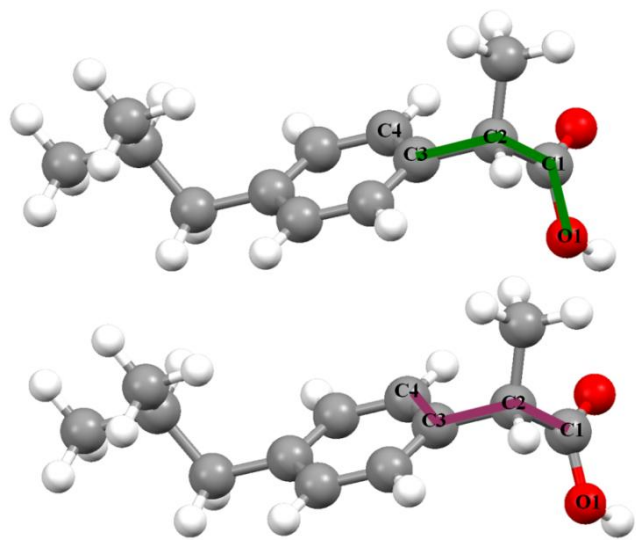

**Fig. 2. Structure and dihedrals defining the conformers in ibuprofen molecule.**

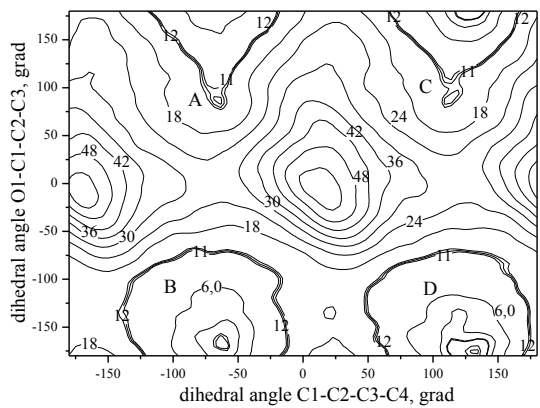

**Fig. 3 2D surface of the free energy for the ibuprofen molecule. Energy values are given in kJ/mol. The dihedral angles are depicted in Figure 2**

Populations of the ibuprofen conformers were calculated based on the free energy values obtained by the MD simulation; they were found to be 0.45 and 0.55 for the conformers A+B and C+D, respectively.

*2.4. Conformer distribution from the NOESY data*

Technical modifications of the high-pressure NMR cell, which were described above, allowed obtaining stable signal with a good enough signal-to-noise ratio and fine spectral resolution. ( see Figure 4) shows that the spectrum resolution is high for all characteristic groups and agrees quite well with what was observed in organic solvents [11]. However, the peaks are broadened as compared with usual liquid-state NMR due to small $T_1$ values in the fluid phase, since the spin-rotational relaxation mechanism in the fluid phase is more effective than in liquids [72].

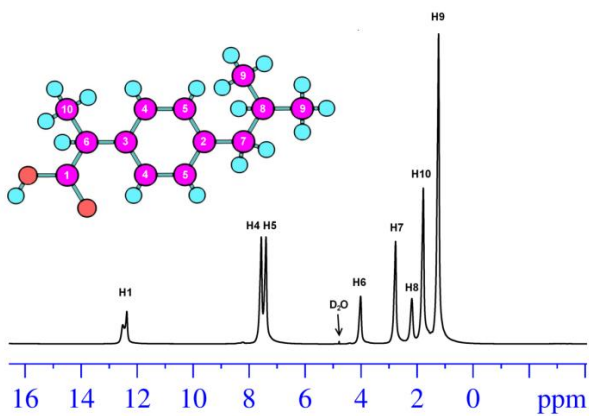

**Fig. 4. $^1$H NMR spectrum of ibuprofen in scCO2 at temperature 50 °C and pressure 130 bar using an deuterium oxide $D_2O$ as external standard in a sealed glass capillary. The chemical structure of ibuprofen and assignments are shown in the figure.**

Two-dimensional nuclear Overhauser effect spectroscopy (NOESY) was used to obtain direct information on the conformer distribution. Cross-relaxation rate observed in NOESY experiments depends strongly on internuclear distances ($\sigma \sim 1/r^6$), and distances up to 5–6 Å can be measured [73].

Cross-relaxation rates obtained from the NOESY spectra are required to reveal conformer populations. In practice, experimental time is limited since it is difficult to maintain stable pressure during several days, and hence we needed to simplify the method of gathering the cross-relaxation rates.

The magnetization transfer rate $d\Delta I_{Zk}/dt_m$ of the nucleus $k$ depends on the spin-lattice relaxation rates $\rho_{kk}$, responsible for the interaction with the lattice, and cross-relaxation rates $\sigma_{kj}$ describing the interaction with other nuclei $j$ of the spin system. In a general case, the magnetization transfer is described by the Solomon equation:

$$\frac{d}{dt_m}\Delta I_{Zk}(t_m) = \rho_{kk}\Delta I_{Zk}(t_m) - \sum_{j \neq k} \sigma_{kj}\Delta I_{Zk}(t_m) \quad (4)$$

This relation applies for each spin in the system, and thus we get a system of differential equations. The magnetization transfer is evaluated by the volumes of all diagonal and all cross-peaks in a 2D NOESY spectrum at a given mixing time $t_m$, which can be gathered in a matrix:

$$A(t_m) = \exp(-Rt_m)A(0) \quad (5)$$

Here $t_m$ is the parameter of the NOESY pulse sequence. The volumes of the peaks in 2D NOESY spectra are usually obtained by numeric integration. The relaxation matrix **R** includes all rates of spin-lattice and cross-relaxations for the system of $n$ spins:

$$R = \begin{pmatrix} \rho_{11} & \sigma_{12} & \cdots & \sigma_{1n} \\ \sigma_{21} & \rho_{22} & \cdots & \sigma_{2n} \\ \vdots & \vdots & \ddots & \vdots \\ \sigma_{n1} & \sigma_{n2} & \cdots & \rho_{nn} \end{pmatrix} \quad (6)$$

The relaxation matrix is calculated accurately enough by fitting Eq. (5) to the experimental peak integrals, measured at different mixing times. It contains all parameters needed to calculate the internuclear distances and thus the molecular structure. Building and analysis of the whole relaxation matrix is a time-consuming task, because a single NOESY spectrum for a small molecule may consist of more than 20 intermolecular dipolar interactions, which correspond to their own distances. Not all these contacts characterise the conformational mobility, and only information on one or two proton pairs involved in the conformational exchange is necessary.

An alternative way to interpret experimental NOESY data assumes that the considered spin system is isolated and can be analyzed alone from the whole spin system [62]. In this case, the cross-peak between two spins, $i$ and $s$, can be expressed as follows [74–77]:

$$A(t_m) = \frac{A_{ii}(0)}{2}\left[1 - \exp(-2\sigma_{is}t_m)\right]\exp\left(\frac{-t_m}{T_{1,is}}\right) \quad (7)$$

where $A_{is}(t_m)$ is the cross-peak volume at the mixing time $t_m$, $A_{ii}(0)$ is the volume of the diagonal peak at $t_m = 0$, and $T_{1,is}$ defines the magnetization losses. Cross-relaxation rates can be derived by analyzing the cross-peak values at different mixing times and volumes of diagonal peaks obtained for a short mixing time. The advantage of the spin pair model is that the cross-peaks corresponding only to those protons which participate in the conformational exchange should be integrated, and there is no need to consider the whole system of magnetization transfer occurring in the molecule. However, it still requires obtaining a series of NOESY spectra with various mixing time. Since recording a single spectrum may long for several tens of hours, the problem of maintaining stable temperature and pressure in the system still remains.

Hence we used a simplified approach allowing determination of the spin-relaxation rate from a single NOESY experiment. In this method the equation for an isolated spin pair is expanded into a Taylor series, and the term linear with the time is taken [78,79]

$$\sigma_{is} = \frac{A_{is}(t_m)}{t_m A_{ii}(t_m)} \quad (8)$$

This approximation is the most effective if spectra are acquired at high temperatures and pressures, because only one NOESY spectrum is required and possible changes of the state parameters is not a problem. Cross-relaxation rates obtained by this method may be biased by centers of magnetization transfer close to the studied spin pair. This possibility was allowed for when choosing the appropriate distances in the molecule.

The strong dependence of the cross-relaxation rates of the distance between the interactions protons is usually approximated by a simple correlation

$$\sigma_{ij} = \frac{n_i \tau_c}{r_i^6} \quad (9)$$

Distance determination is based on the fact that if a calibration distance in the system is known, the unknown distance can be found as [80]

$$r_i = r_{ref} \left( \frac{\sigma_{ref}}{\sigma_i} \right)^{1/6} \quad (10)$$

Lee and Krishna showed that in the case of fast conformational exchange, the cross-relaxation rate observed in a NOESY experiment is an average of the rates corresponding to individual conformers [81]:

$$\sigma_{exp} = \sum_i \sigma_i x_i \quad (11)$$

This equation allows us to express the fractions $x_i$ of the forms in the case of two-position exchange based on the known distances [12]:

$$\frac{1}{r_{exp}^6} = \frac{x_1}{r_1^6} + \frac{1 - x_1}{r_2^6} \quad (12)$$

Here the distances $r_1$ and $r_2$ (in our case, A+B and C+D) are obtained from quantum-chemical calculations reported earlier [14], the population $x_1$ of the first conformers is calculated, and the fraction of the second form is found as $x_2 = 1 - x_1$.

The approach described above was used to determine the fractions of the ibuprofen conformers at the supercritical state parameters. The NOESY spectrum (Figure 5) with the mixing time of 0.3 s contained peaks responsible for the conformation-dependent internuclear distances. The peak H4–OH1 was chosen for this purpose, and the H4–H6 signal was used as the calibration one. The choice of the latter distance is justified by the fact that it is the same in all conformers and gives an intensive NOESY signal, allowing us to integrate the peak with a high accuracy. As was stated in the section of MD simulation, the conformation flexibility of the ibuprofen molecules are due to the mobility of the hydroxyl group with respect to the phenol fragment.

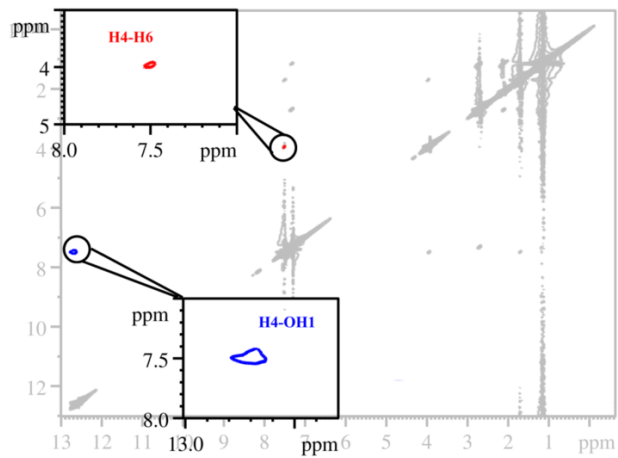

Fig. 5. 2D NOESY spectrum obtained at the supercritical state parameters and the mixing time of 0.3 s. The conformation-dependent distance H4–OH1 and the calibration distance H4–H6 are shown by arrows.

Cross-relaxation rates according to Eq. (8) were found to be 0.00481 s$^{-1}$ for H4–OH1 and 0.06065 s$^{-1}$ for H4–H6. Then the internuclear distance was calculated using Eq. (10); it turned out to be 4.17 Å. Varying the integration limits in the NOESY spectra changes the distance value by 1–3%.

Equation (12) was used then to determine the fractions of the conformers A+B and C+D. Corresponding distances were taken from [82] and were 4.52 and 3.96 Å, respectively. We estimate the standard deviation of finding $r_{exp}$ as ±3%, which agrees to previous reports describing the NOESY analysis [65,68].

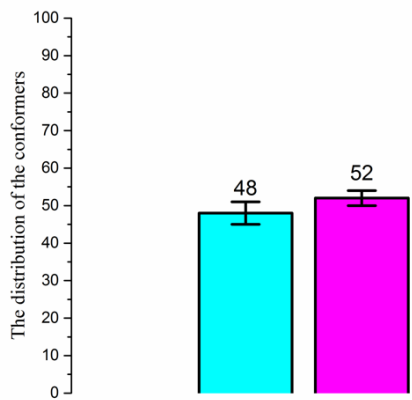

Fig. 6. Conformer distribution for solutions of ibuprofen in supercritical carbon dioxide based on the experimental 2D NOESY data 50 °C and 130 bar . Error in determination of the conformer fractions does not exceed 3%.

Thus, the inaccuracy of finding the conformer populations can be calculated, as it was made for small molecules in [42]. Based on the NOESY data with corresponding errors, the conformers fractions for ibuprofen are 48 and 52% (Figure 6) with the accuracy of ±3%. Varying the values $r_1$ and $r_2$ showed that the errors contained in the quantum-chemical calculations seem to be unimportant. The conformer fractions with the experimental errors are presented in Figure 7. Evidently, within the experimental error the relative fractions of the conformers A+B and C+D is close to 1 : 1.

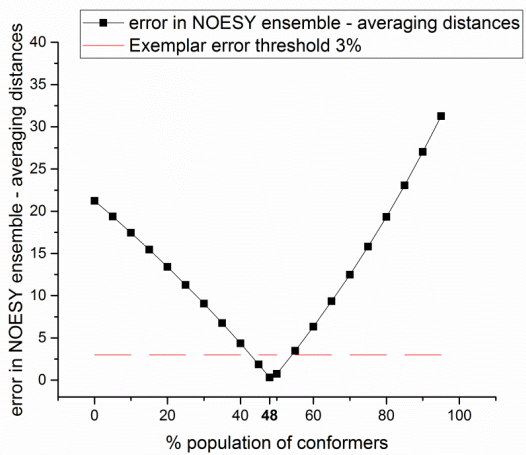

**Fig. 7.** The dependence of relative deviation of interproton distance calculated by eq. 12 from experimental NOESY data on the population of conformers of ibuprofen (left figure). The red dot line corresponds to 3% experimental error in NOESY data.

*2.5. Signal splitting in $^1$H NMR spectrum*

Splitting of the $^1$H NMR signal of the carboxyl group was observed (Figure 7). This is due to the fact that the OH group is subject to the shielding effect from the benzene ring of ibuprofen to a different extent in different conformers. When the OH group is located near the protons of the benzene ring, its signal is observed in a higher field. Indeed, the hydroxyl group in forms C and D is at 3.48 Å from the nearest proton of the benzene ring, which is almost 1 Å smaller than what is found in the conformers A and B.

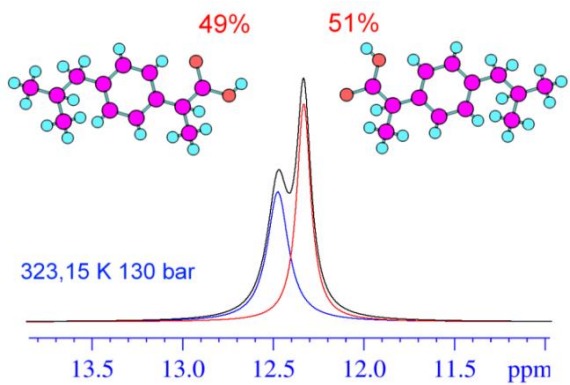

**Fig. 7.** Part of $^1$H NMR spectrum of ibuprofen in scCO$_2$ at 50°C and 130 bar. Error in determination of the conformer fractions does not exceed 5%.

This effect is of the critical nature, though it is not directly related to the supercritical parameters of state (temperature and pressure). In the NMR experiment we deal with a supersaturated solution of ibuprofen in supercritical carbon dioxide. In fact, we study interface with clearly separated bottom and bulk. Despite the fact that the bottom phase is outside the sensitive region of the probe, it affects the observed values. The chemical potentials of the phases are equal, and the observed conformational distribution in the bulk is permanently nurtured which is observed as slowing down the conformational exchange rate.

The signal of the OH group was decomposed into components to determine the conformational composition of ibuprofen in scCO$_2$. A similar procedure used in processing IR spectra is described in [83,84]. As a result, the integral intensities of the components of the OH signal were found to be 49 : 51, which is in a good agreement with the values obtained from the MD simulation and NOESY spectrum at the same parameters of state (Table 1).

**Table 1.** Distribution of the conformer populations of ibuprofen. Experimental error level for the NMR and NOESY results is about ±5% and ±3% correspondently.

| Conformers | MD | NOESY | NMR |
|---|---|---|---|
| A+B | 0.45 | 0.48 | 0.49 |
| C+D | 0.55 | 0.52 | 0.51 |

## 3. Conclusion

We modified the equipment for conducting NMR measurements at the supercritical parameters of state, which allowed not only to increase the sensitivity of the NMR experiments. The method of estimating the conformational state of small molecules based on a single NOESY measurement was proposed and tested. The presence of the splitting of OH group NMR signal into two lines most likely be the result of conformational exchange. Furthermore, a conformational equilibrium of ibuprofen in scCO$_2$ was observed with the conformer fractions of 49% and 51%, which agree well with the MD simulation results.

## 4. Experimental Section

Ibuprofen was purchased from Aldrich (purity ≥99.5%); high-purity carbon dioxide (CO$_2$ 99.995%, H$_2$O <0.001%) was purchased from Linde Gas (the Linde Gas Group, Balashikha, Russia). The chemical were used without further purification. For all high-pressure NMR experiments, the samples were transferred into a single-crystal sapphire tube and dissolved in CO$_2$. To minimize the influence of ambient air paramagnetic oxygen on the sample, we purged sapphire tube with the pure CO$_2$ The experiments were carried out using a home-built universal experimental setup [85]. Equilibration time was set as 40 min after applying pressure and temperature changes. A defined CO$_2$ pressure was adjusted using manual press. The temperature was controlled by a BVT-2000 unit and Bruker cooling unit BCU, gas flow was 535 l/h. The temperature calibration was carried out using a standard K-type thermocouple as a reference. As a verification method for temperature calibration we used the $^1$H NMR signal of methanol[86].

High-pressure NMR experiments were performed on a Bruker Avance III 500 instrument equipped with a Bruker 5-mm TBI probe. $^1$H NMR spectra were recorded at 50 MHz using π/2 pulses, relaxation delay of 2 s, and spectral width of 14 ppm. A sealed capillary with D$_2$O was used for field stabilization. The peaks in the high-pressure spectra were referenced using the high-resolution spectra in chloroform. The dcon program in Topspin 2.1 was used to carry out the processing and the fitting of the NMR $^1$H spectrum. The Lorentzian shape was chosen for the approximation procedure, but the Gaussian weighting function was introduced to allow for possible inhomogeneity of the magnetic field.

The 2D NOESY experiment was performed with 2048 × 512 data. The relaxation delay was chosen to be 2.5 s, and the π/2 pulse length was 9.3 μs. The value of mixing time was taken as 300 ms. The spectral processing for NOESY was made in Bruker Topspin 2.1 software. The resulting multiple FID for the 2D NOESY experiment was filtered with the sine-squared function and in both dimensions before performing Fourier transformation.

### Acknowledgements

This work was supported by the Russian Foundation for Basic Research (grants No. 18-29-06008 and 18-03-00255); by Ministry of Education and Science of the Russian Federation (contract No. 01201260481 and 0120095082) and under grant of Council on grants of

REFERENCES

[1] N. Chieng, T. Rades, J. Aaltonen, An overview of recent studies on the analysis of pharmaceutical polymorphs, J. Pharm. Biomed. Anal. 55 (2011) 618–644. doi:10.1016/j.jpba.2010.12.020.

[2] S.L. Morissette, Ö. Almarsson, M.L. Peterson, J.F. Remenar, M.J. Read, A. V. Lemmo, S. Ellis, M.J. Cima, C.R. Gardner, High-throughput crystallization: Polymorphs, salts, co-crystals and solvates of pharmaceutical solids, Adv. Drug Deliv. Rev. 56 (2004) 275–300. doi:10.1016/j.addr.2003.10.020.

[3] S.L. Price, From crystal structure prediction to polymorph prediction: interpreting the crystal energy landscape, Phys. Chem. Chem. Phys. 10 (2008) 1996. doi:10.1039/b719351c.

[4] M. Pudipeddi, A.T.M. Serajuddin, Trends in solubility of polymorphs, J. Pharm. Sci. 94 (2005). doi:10.1002/jps.20302.

[5] G.P. Stahly, Diversity in single- and multiple-component crystals. the search for and prevalence of polymorphs and cocrystals, Cryst. Growth Des. 7 (2007). doi:10.1021/cg060838j.

[6] I. Weissbuch, L. Addadi, M. Lahav, L. Leiserowitz, Molecular recognition at crystal interfaces, Science (80-. ). 253 (1991).

[7] P.A. Kalmykov, I.A. Khodov, V.V. Klochkov, M.V. Klyuev, Theoretical and experimental study of imine-enamine tautomerism of condensation products of propanal with 4-aminobenzoic acid in ethanol, Russ. Chem. Bull. 66 (2017) 70–75. doi:10.1007/s11172-017-1701-3.

[8] S.I. Selivanov, A.G. Shavva, An NMR study of the spatial structure and intramolecular dynamics of modified analogues of steroid hormones, Russ. J. Bioorganic Chem. 28 (2002). doi:10.1023/A:1015704203799.

[9] C.P. Butts, C.R. Jones, Z. Song, T.J. Simpson, Accurate NOE-distance determination enables the stereochemical assignment of a flexible molecule - arugosin C, Chem. Commun. 48 (2012) 9023–9025. doi:10.1039/C2CC32144K.

[10] I.A. Khodov, M.Y. Nikiforov, G.A. Alper, D.S. Blokhin, S.V. Efimov, V.V. Klochkov, N. Georgi, Spatial structure of felodipine dissolved in DMSO by 1D NOE and 2D NOESY NMR spectroscopy, J. Mol. Struct. 1035 (2013) 358–362. doi:10.1016/j.molstruc.2012.11.040.

[11] I.A. Khodov, S.V. Efimov, V.V. Klochkov, G.A. Alper, L.A.E. Batista De Carvalho, Determination of preferred conformations of ibuprofen in chloroform by 2D NOE spectroscopy., Eur. J. Pharm. Sci. 65C (2014) 65–73. doi:10.1016/j.ejps.2014.08.005.

[12] I.A. Khodov, S. V. Efimov, M.Y. Nikiforov, V. V. Klochkov, N. Georgi, Inversion of population distribution of felodipine conformations at increased concentration in dimethyl sulfoxide is a prerequisite to crystal nucleation., J. Pharm. Sci. 103 (2014) 392–394. doi:10.1002/jps.23833.

[13] A. Kolmer, L.J. Edwards, I. Kuprov, C.M. Thiele, Conformational analysis of small organic molecules using NOE and RDC data: A discussion of strychnine and a -methylene- c -butyrolactone, J. Magn. Reson. 261 (2015) 101–109. doi:10.1016/j.jmr.2015.10.007.

[14] I.A. Khodov, S. V. Efimov, V. V. Klochkov, L.A.E. Batista De Carvalho, M.G. Kiselev, The importance of suppressing spin diffusion effects in the accurate determination of the spatial structure of a flexible molecule by nuclear Overhauser effect spectroscopy, J. Mol. Struct. 1106 (2016) 373–381. doi:10.1016/j.molstruc.2015.10.055.

[15] S.V. Efimov, I.A. Khodov, E.L. Ratkova, M.G. Kiselev, S. Berger, V.V. Klochkov, Detailed NOESY/T-ROESY analysis as an effective method for eliminating spin diffusion from 2D NOE spectra of small flexible molecules, J. Mol. Struct. 1104 (2016) 63–69. doi:10.1016/j.molstruc.2015.09.036.

[16] M.E. Di Pietro, G. Celebre, C. Aroulanda, D. Merlet, G. De Luca, Assessing the stable conformations of ibuprofen in solution by means of Residual Dipolar Couplings, Eur. J. Pharm. Sci. 106 (2017). doi:10.1016/j.ejps.2017.05.029.

[17] E.S. Machado, D.A. Silva, K.J. de Almeida, V.C. Felicíssimo, Conformational analysis and vibrational spectroscopic studies of tetraethoxysilane and its hydrolysis products: A DFT prediction, J. Mol. Struct. 1134 (2017) 360–368. doi:10.1016/j.molstruc.2016.12.104.

[18] K. Subashini, S. Periandy, Spectroscopic (FT-IR, FT-Raman, UV, NMR, NLO) investigation and molecular docking study of 1-(4-Methylbenzyl) piperazine, J. Mol. Struct. 1134 (2017) 157–170. doi:10.1016/j.molstruc.2016.12.048.

[19] S. Premkumar, T.N. Rekha, R. Mohamed Asath, T. Mathavan, A. Milton Franklin Benial, Vibrational spectroscopic, molecular docking and density functional theory studies on 2-acetylamino-5-bromo-6-methylpyridine, Eur. J. Pharm. Sci. 82 (2016). doi:10.1016/j.ejps.2015.11.018.

[20] R.D. Oparin, D.V. Ivlev, A.M. Vorobei, A. Idrissi, M.G. Kiselev, Screening of conformational polymorphism of ibuprofen in supercritical $CO_2$, J. Mol. Liq. 239 (2017) 49–60. doi:10.1016/j.molliq.2016.10.132.

[21] R.D. Oparin, M. Moreau, I. De Walle, M. Paolantoni, A. Idrissi, M.G. Kiselev, The interplay between the paracetamol polymorphism and its molecular structures dissolved in supercritical $CO_2$ in contact with the solid phase: In situ vibration spectroscopy and molecular dynamics simulation analysis, Eur. J. Pharm. Sci. 77 (2015) 48–59. doi:10.1016/j.ejps.2015.05.016.

[22] R.D. Oparin, D.V. Ivlev, M.G. Kiselev, Conformational equilibria of pharmaceuticals in supercritical $CO_2$, IR spectroscopy and quantum chemical calculations, Spectrochim. Acta Part A Mol. Biomol. Spectrosc. 230 (2020) 118072. doi:10.1016/J.SAA.2020.118072.

[23] R.D. Oparin, M. V. Kurskaya, M.A. Krestyaninov, A. Idrissi, M.G. Kiselev, Correlation between the conformational crossover of carbamazepine and its polymorphic transition in supercritical $CO_2$: On the way to polymorph control, Eur. J. Pharm. Sci. (2020) 105273. doi:10.1016/J.EJPS.2020.105273.

[24] T. Tsukahara, Y. Kayaki, T. Ikariya, Y. Ikeda, 13C NMR spectroscopic evaluation of the affinity of carbonyl compounds for carbon dioxide under supercritical conditions, Angew. Chemie - Int. Ed. 43 (2004). doi:10.1002/anie.200454190.

[25] B. Chandrika, L.K. Schnackenberg, P. Raveendran, S.L. Wallen, High resolution 1H NMR structural studies of sucrose octaacetate in supercritical carbon dioxide, Chem. - A Eur. J. 11 (2005). doi:10.1002/chem.200500215.

[26] G. Uccello-Barretta, G. Sicoli, F. Balzano, P. Salvadori, A conformational model of per-O-acetyl-cyclomaltoheptaose (-β-cyclodextrin) in solution: Detection of partial inversion of glucopyranose units by NMR spectroscopy, Carbohydr. Res. 338 (2003). doi:10.1016/S0008-6215(03)00074-0.

[27] G.I. Ivanova, E.R. Vão, M. Temtem, A. Aguiar-Ricardo, T. Casimiro, E.J. Cabrita, High-pressure NMR characterization of triacetyl-β-cyclodextrin in supercritical carbon dioxide, Magn. Reson. Chem. 47 (2009). doi:10.1002/mrc.2365.

[28] H. C. Hoffmann, B. Assfour, F. Epperlein, N. Klein, S. Paasch, I. Senkovska, S. Kaskel, G. Seifert, E. Brunner, High-Pressure in Situ 129Xe NMR Spectroscopy and Computer Simulations of Breathing Transitions in the Metal–Organic Framework Ni2(2,6-ndc)2(dabco) (DUT-8(Ni)), J. Am. Chem. Soc. 133 (2011) 8681–8690. doi:10.1021/ja201951t.

[29] F. Kolbe, S. Krause, V. Bon, I. Senkovska, S. Kaskel, E. Brunner, High-Pressure in Situ 129Xe NMR Spectroscopy: Insights into Switching Mechanisms of Flexible Metal–Organic Frameworks Isoreticular to DUT-49, Chem. Mater. 31 (2019) 6193–6201. doi:10.1021/acs.chemmater.9b02003.

[30] M.C. Corvo, J. Sardinha, S.C. Menezes, S. Einloft, M. Seferin, J. Dupont, T. Casimiro, E.J. Cabrita, Solvation of Carbon Dioxide in [C4mim][BF4] and [C4mim][PF6] Ionic Liquids Revealed by High-Pressure NMR Spectroscopy, Angew. Chemie Int. Ed. 52 (2013) 13024–13027. doi:10.1002/anie.201305630.

[31] N.M. Simon, M. Zanatta, F.P. dos Santos, M.C. Corvo, E.J. Cabrita, J. Dupont, Carbon Dioxide Capture by Aqueous Ionic Liquid Solutions, ChemSusChem. 10 (2017) 4927–4933. doi:10.1002/cssc.201701044.

[32] M.C. Corvo, J. Sardinha, T. Casimiro, G. Marin, M. Seferin, S. Einloft, S.C. Menezes, J. Dupont, E.J. Cabrita, A Rational Approach to CO2 Capture by Imidazolium Ionic Liquids: Tuning CO2 Solubility by Cation Alkyl Branching, ChemSusChem. 8 (2015) 1935–1946. doi:10.1002/cssc.201500104.

[33] C.D. Pilgrim, C.A. Colla, G. Ochoa, J.H. Walton, W.H. Casey, 29Si NMR of aqueous silicate complexes at gigapascal pressures, Commun. Chem. 1 (2018) 67. doi:10.1038/s42004-018-0066-3.

[34] C.A. Ohlin, W.H. Casey, 17O NMR as a Tool in Discrete Metal Oxide Cluster Chemistry, Annu. Reports NMR Spectrosc. 94 (2018) 187–248. doi:10.1016/BS.ARNMR.2018.01.001.

[35] G. Ochoa, C.A. Colla, P. Klavins, M.P. Augustine, W.H. Casey, NMR spectroscopy of some electrolyte solutions to 1.9 GPa, Geochim. Cosmochim. Acta. 193 (2016) 66–74. doi:10.1016/J.GCA.2016.08.013.

[36] G. Ochoa, C.D. Pilgrim, M.N. Martin, C.A. Colla, P. Klavins, M.P. Augustine, W.H. Casey, 2H and 139La NMR Spectroscopy in Aqueous Solutions at Geochemical Pressures, Angew. Chemie Int. Ed. 54 (2015) 15444–15447. doi:10.1002/anie.201507773.

[37] E.D. Walter, L. Qi, A. Chamas, H.S. Mehta, J.A. Sears, S.L. Scott, D.W. Hoyt, Operando MAS NMR Reaction Studies at High Temperatures and Pressures, J. Phys. Chem. C. 122 (2018) 8209–8215. doi:10.1021/acs.jpcc.7b11442.

[38] I.Z. Rakhmatullin, L.F. Galiullina, M.R. Garipov, A.D. Strel'nik, Y.G. Shtyrlin, V. V Klochkov, Dynamic NMR study of dinitrophenyl derivatives of seven-membered cyclic ketals of pyridoxine., Magn. Reson. Chem. 53 (2015) 805–12. doi:10.1002/mrc.4251.

[39] I.Z. Rakhmatullin, L.F. Galiullina, E.A. Klochkova, I.A. Latfullin, A.V. Aganov, V.V. Klochkov, Structural studies of pravastatin and simvastatin and their complexes with SDS micelles by NMR spectroscopy, J. Mol. Struct. 1105 (2016) 25–29. doi:10.1016/j.molstruc.2015.10.059.

[40] R.M. Aminova, L.F. Galiullina, N.I. Silkin, A.R. Ulmetov, V.V. Klochkov, A.V. Aganov, Investigation of complex formation between hydroxyapatite and fragments of collagen by NMR spectroscopy and quantum-chemical modeling, J. Mol. Struct. 1049 (2013) 13–21. doi:10.1016/j.molstruc.2013.06.008.

[41] S.I. Selivanov, S. Wang, A.S. Filatov, A. V Stepakov, NMR Study of Spatial Structure and Internal Dynamic of Adducts of Ninhydrin-Derived Azomethine Ylide with Cyclopropenes, Appl. Magn. Reson. (2019). doi:10.1007/s00723-019-01178-w.

[42] G.A. Gamov, I.A. Khodov, K.V. Belov, M.N. Zavalishin, A.N. Kiselev, T.R. Usacheva, V.A. Sharnin, Spatial structure, thermodynamics and kinetics of formation of hydrazones derived from pyridoxal 5′-phosphate and 2-furoic, thiophene-2-carboxylic hydrazides in solution, J. Mol. Liq. 283 (2019) 825–833. doi:10.1016/J.MOLLIQ.2019.03.125.

[43] G.B. Benedek, E.M. Purcell, Nuclear Magnetic Resonance in Liquids under High Pressure, J. Chem. Phys. 22 (1954) 2003–2012. doi:10.1063/1.1739982.

[44] J. Jonas, L. Ballard, D. Nash, High-Resolution, High-Pressure NMR Studies of Proteins, Biophys. J. 75 (1998) 445–452. doi:10.1016/S0006-3495(98)77532-0.

[45] L. Ballard, C. Reiner, J. Jonas, High-Resolution NMR Probe for Experiments at High Pressures, J. Magn. Reson. Ser. A. 123 (1996) 81–86. doi:10.1006/jmra.1996.0216.

[46] B.G. Pautler, C.A. Colla, R.L. Johnson, P. Klavins, S.J. Harley, C.A. Ohlin, D.A. Sverjensky, J.H. Walton, W.H. Casey, A High-Pressure NMR Probe for Aqueous Geochemistry, Angew. Chemie Int. Ed. 53 (2014) 9788–9791. doi:10.1002/anie.201404994.

[47] H.S. Mehta, Y. Chen, J.A. Sears, E.D. Walter, M. Campos, J. Kothandaraman, D.J. Heldebrant, D.W. Hoyt, K.T. Mueller, N.M. Washton, A novel high-temperature MAS probe with optimized temperature gradient across sample rotor for in-situ monitoring of high-temperature high-pressure chemical reactions, Solid State Nucl. Magn. Reson. 102 (2019) 31–35. doi:10.1016/J.SSNMR.2019.06.003.

[48] H. Yamada, K. Nishikawa, M. Honda, T. Shimura, K. Akasaka, K. Tabayashi, Pressure-resisting cell for high-pressure, high-resolution nuclear magnetic resonance measurements at very high magnetic fields, Rev. Sci. Instrum. 72 (2001) 1463–1471. doi:10.1063/1.1334630.

[49] D. Christopher Roe, Sapphire NMR tube for high-resolution studies at elevated pressure, J. Magn. Reson. 63 (1985) 388–391. doi:10.1016/0022-2364(85)90332-4.

[50] J.L. Urbauer, M.R. Ehrhardt, R.J. Bieber, P.F. Flynn, a. J. Wand, High-resolution triple-resonance NMR spectroscopy of a novel calmodulin peptide complex at kilobar pressures, J. Am. Chem. Soc. 118 (1996) 11329–11330. doi:10.1021/ja962552g.

[51] J. Roche, C.A. Royer, C. Roumestand, Monitoring protein folding through high pressure NMR spectroscopy, Prog. Nucl. Magn. Reson. Spectrosc. 102-103 (2017) 15–31. doi:10.1016/J.PNMRS.2017.05.003.

[52] P. Hirunsit, Z. Huang, T. Srinophakun, M. Charoenchaitrakool, S. Kawi, Particle formation of ibuprofen-supercritical CO$_2$ system from rapid expansion of supercritical solutions (RESS): A mathematical model, Powder Technol. 154 (2005). doi:10.1016/j.powtec.2005.03.020.

[53] M. Türk, G. Upper, M. Steurenthaler, K. Hussein, M.A. Wahl, Complex formation of Ibuprofen and β-Cyclodextrin by controlled particle deposition (CPD) using SC-CO$_2$, J. Supercrit. Fluids. 39 (2007). doi:10.1016/j.supflu.2006.02.009.

[54] M. Muntó, N. Ventosa, S. Sala, J. Veciana, Solubility behaviors of ibuprofen and naproxen drugs in liquid "CO2-organic solvent" mixtures, J. Supercrit. Fluids. 47 (2008). doi:10.1016/j.supflu.2008.07.013.

[55] D. Velasco, L. Benito, M. Fernández-Gutiérrez, J. San Román, C. Elvira, Preparation in supercritical CO$_2$ of porous poly(methyl methacrylate)-poly(l-lactic acid) (PMMA-PLA) scaffolds incorporating ibuprofen, J. Supercrit. Fluids. 54 (2010). doi:10.1016/j.supflu.2010.05.012.

[56] P.W. Labuschagne, S.G. Kazarian, R.E. Sadiku, Supercritical CO2-assisted preparation of ibuprofen-loaded PEG-PVP complexes, J. Supercrit. Fluids. 57 (2011). doi:10.1016/j.supflu.2011.03.001.

[57] G. Dannhardt, W. Kiefer, Cyclooxygenase inhibitors – current status and future prospects, Eur. J. Med. Chem. 36 (2001) 109–126. doi:10.1016/S0223-5234(01)01197-7.

[58] M.L. Vueba, M.E. Pina, L.A.E.B. De Carvalho, Conformational Stability of Ibuprofen: Assessed by DFT Calculations and Optical Vibrational Spectroscopy, J. Pharm. Sci. 97 (2008) 845–859. doi:10.1002/jps.

[59] E. Dudognon, F. Danède, Evidence for a new crystalline phase of racemic ibuprofen, Pharm. Res. 25 (2008) 2853–2858. doi:10.1007/s11095-008-9655-7.

[60] J. P. M. Jämbeck, A. P. Lyubartsev, Exploring the Free Energy Landscape of Solutes Embedded in Lipid Bilayers, J. Phys. Chem. Lett. 4 (2013) 1781–1787. doi:10.1021/jz4007993.

[61] S. M. Loverde, Molecular Simulation of the Transport of Drugs across Model Membranes, J. Phys. Chem. Lett. 5 (2014) 1659–1665. doi:10.1021/jz500321d.

[62] M. Geppi, S. Guccione, G. Mollica, R. Pignatello, C.A. Veracini, Molecular Properties of Ibuprofen and Its Solid Dispersions with Eudragit RL100 Studied by Solid-State Nuclear Magnetic Resonance, Pharm. Res. 22 (2005) 1544–1555. doi:10.1007/s11095-005-6249-5.

[63] A. Laio, M. Parrinello, Escaping free-energy minima, Proc. Natl. Acad. Sci. U. S. A. 99 (2002). doi:10.1073/pnas.202427399.

[64] A. Laio, F.L. Gervasio, Metadynamics: A method to simulate rare events and reconstruct the free energy in biophysics, chemistry and material science, Reports Prog. Phys. 71 (2008). doi:10.1088/0034-4885/71/12/126601.

[65] M. Bonomi, D. Branduardi, G. Bussi, C. Camilloni, D. Provasi, P. Raiteri, D. Donadio, F. Marinelli, F. Pietrucci, R.A. Broglia, M. Parrinello, PLUMED: A portable plugin for free-energy calculations with molecular dynamics, Comput. Phys. Commun. 180 (2009). doi:10.1016/j.cpc.2009.05.011.

[66] I.V. Fedorova, D.V. Ivlev, M.G. Kiselev, Conformational lability of ibuprofen in supercritical carbon dioxide, Russ. J. Phys. Chem. B. 10 (2016) 1153–1162. doi:10.1134/S199079311607006X.

[67] J. Wang, R.M. Wolf, J.W. Caldwell, P.A. Kollman, D.A. Case, Development and testing of a general Amber force field, J. Comput. Chem. 25 (2004). doi:10.1002/jcc.20035.

[68] L. Kalé, R. Skeel, M. Bhandarkar, R. Brunner, A. Gursoy, N. Krawetz, J. Phillips, A. Shinozaki, K. Varadarajan, K. Schulten, NAMD2: Greater Scalability for Parallel Molecular Dynamics, J. Comput. Phys. 151 (1999) 283–312. doi:https://doi.org/10.1006/jcph.1999.6201.

[69] P.J. Linstrom, W.G. Mallard, The NIST Chemistry WebBook: A chemical data resource on the Internet, J. Chem. Eng. Data. 46 (2001). doi:10.1021/je000236i.

[70] L. Martinez, R. Andrade, E.G. Birgin, J.M. Martínez, PACKMOL: A package for building initial configurations for molecular dynamics simulations, J. Comput. Chem. 30 (2009). doi:10.1002/jcc.21224.

[71] Z. Zhang, Z. Duan, An optimized molecular potential for carbon dioxide, J. Chem. Phys. 122 (2005). doi:10.1063/1.1924700.

[72] A. V Aganov, V. V Klochkov, Y.Y. Samitov, New aspects of the application of nuclear magnetic resonance to the study of chemical exchange processes, Russ. Chem. Rev. 54 (1985) 1585–1612. doi:10.1070/RC1985v054n10ABEH003123.

[73] M. Madrid, E. Llinás, M. Llinás, Model-independent refinement of interproton distances generated from 1H NMR overhauser intensities, J. Magn. Reson. 93 (1991) 329–346.

[74] S. Macura, R.R. Ernst, Elucidation of cross relaxation in liquids by two-dimensional N.M.R. spectroscopy, Mol. Phys. 100 (2002) 135–147. doi:10.1080/00268970110088983.

[75] H.A. Scheidt, D. Huster, The interaction of small molecules with phospholipid membranes studied by 1 H NOESY NMR under magic-angle spinning, Acta Pharmacol. Sin. 29 (2008) 35.

[76] H.A. Scheidt, R.M. Badeau, D. Huster, Investigating the membrane orientation and transversal distribution of 17β-estradiol in lipid membranes by solid-state NMR, Chem. Phys. Lipids. 163 (2010) 356–361. doi:10.1016/J.CHEMPHYSLIP.2010.02.001.

[77] L.E. Nikitina, R.S. Pavelyev, V.A. Startseva, S. V. Kiselev, L.F. Galiullina, O. V. Aganova, A.F. Timerova, S. V. Boichuk, Z.R. Azizova, V. V. Klochkov, D. Huster, I.A. Khodov, H.A. Scheidt, Structural details on the interaction of biologically active sulfur-containing monoterpenoids with lipid membranes, J. Mol. Liq. 301 (2020) 112366. doi:10.1016/J.MOLLIQ.2019.112366.

[78] H. Hu, K. Krishnamurthy, Revisiting the initial rate approximation in kinetic NOE measurements., J. Magn. Reson. 182 (2006) 173–177. doi:10.1016/j.jmr.2006.06.009.

[79] D. Huster, K. Arnold, K. Gawrisch, Investigation of Lipid Organization in Biological Membranes by Two-Dimensional Nuclear Overhauser Enhancement Spectroscopy, J. Phys. Chem. B. 103 (1998) 243–251. doi:10.1021/jp983428h.

[80] I.A. Khodov, M.G. Kiselev, S.V. Efimov, V.V. Klochkov, Comment on "Conformational analysis of small organic molecules using NOE and RDC data: A discussion of strychnine and α-methylene-γ-butyrolactone"., J. Magn. Reson. 266 (2016). doi:10.1016/j.jmr.2016.02.009.

[81] W. Lee, N. Krishna, Influence of conformational exchange on the 2D NOESY spectra of biomolecules existing in multiple conformations, J. Magn. Reson. 98 (1992) 36–48. doi:10.1016/0022-2364(92)90107-I.

[82] O. Maltceva, G. Mamardashvili, I. Khodov, D. Lazovskiy, V. Khodova, M. Krest'yaninov, N. Mamardashvili, W. Dehaen, Molecular recognition of nitrogen – containing bases by Zn[5,15-bis-(2,6-dodecyloxyphenyl)]porphyrin, Supramol. Chem. 29 (2017) 360–369. doi:10.1080/10610278.2016.1238473.

[83] A.A. Dyshin, O.V. Eliseeva, G.V. Bondarenko, A.M. Kolker, M.G. Kiselev, Dispersion of single-walled carbon nanotubes in dimethylacetamide and a dimethylacetamide–cholic acid mixture, Russ. J. Phys. Chem. A. 90 (2016). doi:10.1134/S0036024416120086.

[84] A.A. Dyshin, R.D. Oparin, M.G. Kiselev, Order structure of methanol along the 200-bar isobar in the temperature range of 60-320°C according to ir spectroscopy, Russ. J. Phys. Chem. B. 6 (2012). doi:10.1134/S1990793112080106.

[85] J. Bonjoch, D. Solé, Synthesis of strychnine, Chem. Rev. 100 (2000) 3455–3482. doi:10.1021/cr9902547.

[86] D.S. Raiford, C.L. Fisk, E.D. Becker, Calibration of methanol and ethylene glycol nuclear magnetic resonance thermometers, Anal. Chem. 51 (1979) 2050–2051. doi:10.1021/ac50048a040.